# A Review of CUDA, MapReduce, and Pthreads Parallel Computing Models

Kato Mivule[1], Benjamin Harvey[2], Crystal Cobb[3], and Hoda El Sayed[4]

[1] Computer Science Department, Bowie State University,
Bowie, MD, USA
kmivule@gmail.com
[2]benharvey20@hotmail.com
[3]misscrys222@gmail.com
[4]helsayed@bowiestate.edu

**Abstract** – *The advent of high performance computing (HPC) and graphics processing units (GPU), present an enormous computation resource for Large data transactions (big data) that require parallel processing for robust and prompt data analysis. While a number of HPC frameworks have been proposed, parallel programming models present a number of challenges – for instance, how to fully utilize features in the different programming models to implement and manage parallelism via multi-threading in both CPUs and GPUs. In this paper, we take an overview of three parallel programming models, CUDA, MapReduce, and Pthreads. The goal is to explore literature on the subject and provide a high level view of the features presented in the programming models to assist high performance users with a concise understanding of parallel programming concepts and thus faster implementation of big data projects using high performance computing.*

## 1. INTRODUCTION

The increasing volume of data (big data) generated by entities unquestionably, require high performance parallel processing models for robust and speedy data analysis. The need for parallel computing has resulted in a number of programming models proposed for high performance computing. However, parallel programming models that interface between the high performance machines and programmers, present challenges; with one of the difficulties being, how to fully utilize features in the different programming models to implement computational units, such as, multi-threads, on both CPUs and GPUs efficiently. Yet still, with the advent of GPUs, additional computation resources for the big data challenge are presented. However, fully utilizing GPU parallelization resources in translating sequential code for HPC implementation is a challenge. Therefore, in this paper, we take an overview of three parallel programming models, CUDA, MapReduce, and Pthreads. Our goal in this study is to give an overall high level view of the features presented in the parallel programming models to assist high performance computing users with a faster understanding of parallel programming concepts and thus better implementation of big data parallel programming projects. Although a number of models could have been chosen for this overview, we focused on CUDA, MapReduce, and Pthreads because of the underlying features in the three models that could be used to for multi-threading. The rest of this paper is organized as follows: In Section 2, a review of the latest literature on parallel computing models and determinism is done. In Section 3, an overview of CUDA is given, while in Section a review of MapReduce features is presented. In Section 5, a topographical exploration of features in Pthreads is given. Finally, in Section 6, a conclusion is done.

## 2. BACKGROUND AND RELATED WORK

As HPC systems increasingly becomes a necessity in mainstream computing, a number of researchers have done work on documenting various features in parallel computing models. Although a number of features in parallel programming models are discussed in literature, CUDA, MapReduce, and Pthreads models, utilize multi-threading and, as such, a look at abstraction and determinism in multi-threading, is given consideration in this paper. In their survey on the parallel programming models, Kasim, March, Zhang, and See (2008), described how parallelism is abstracted and presented to programmers, by investigating six parallel programming models used by the HPC community, namely, Pthreads, OpenMP, CUDA, MPI, UPC, and Fortress [1]. Furthermore, Kasim et al, proposed a set of criteria to evaluate how parallelism is abstracted and presented to programmers, namely, (i) system architecture, (ii) programming methodologies, (iii) worker management, (iv) workload partitioning scheme, (v) task-to-worker mapping, (vi) synchronization, and (vii) communication model [1].

On the feature of determinism in parallel programming, Bocchino, Adve, Adve, and Snir

(2009), noted that with most parallel programming models, subtle coding errors could lead to inadvertent non-deterministic actions and bugs that are hard to find and debug [2]. Therefore, Bocchino et al, argued for a parallel programming model that was deterministic by default unless the programmer unambiguously decided to use non-deterministic constructs [2]. Non-determinism is a condition in multi-threading parallel processing, in which erratic and unpredictable behavior, bugs, and errors occur among threads due to their random interleaving and implicit communication when accessing shared memory, thus making it more problematic to program in parallel model than the sequential von Neumann model [3] [4].

Additionally, on the feature of determinism in parallel programming languages, McCool (2010) argued that a more structured approach in the design and implementation of parallel algorithms was needed in order to reduce the complexity in developing software [5]. Moreover, McCool (2010) suggested the utilization of deterministic algorithmic skeletons and patterns in the design of parallel programs [5]. McCool (2010) generally proposed that any system that maintains a specification and composition of a high-quality pattern can be used as a template to steer developers in designing more dependable deterministic parallel programs [5]. However, Yang, Cui, Wu, Tang, and Hu (2013) contended that determinism in parallel programming was difficult to achieve since threads are non-deterministic by default and therefore making parallel programs reliable with stable multi-threading was more practical [6]. On the other hand, Garland, Kudlur, and Zheng (2012), observed that while heterogeneous architectures and massive multi-threading concepts have been accepted in the larger HPC community, modern programming models suffer from a defect of limited concurrency, undifferentiated flat memory storage, and a homogenous processing of elements [7]. Garland et al, then, proposed a unified programming model for heterogeneous machines that provides constructs for massive parallelism, synchronization, and data assignment, executed on the whole machine [7]. Dean and Ghemawat [33] presented a run-time parallel, distributed, scalable programming model, which encompassed a map and reduce function in order to map and reduce key-value pairs [33].

Though there were many parallel, distributed, scalable programming models that currently existed, there was no model in existence whose goal through development was to create a programming model that was easy to use. The innate ability of MapReduce to do its parallel and distributed computation across large commodity clusters at run-time allowed developers to express simple computations while hiding the messy details of communication, parallelization, fault tolerance, and load balancing [33]. On fully utilizing the GPU and MapReduce model, Ji and Ma (2011), explored the prospective value of allowing for a GPU MapReduce structure that uses multiple levels of the GPU memory hierarchy [8]. To achieve this goal, Ji and Ma (2011) proposed a resourceful deployment using small shared but fast memory, which included, a shared memory staging area management, thread-role partitioning, and intra-block thread synchronization [8].

Furthermore, Fang, He, Luo, and Govindaraju (2011) proposed the Mars model, which combines GPU and MapReduce features to attain: (i) ease of programming GPUs, (ii) portability such that the system can be applied on different technologies such as NVIDIA CUDA, or Pthreads, and (iii) effective use of high performance using GPUs [9]. To achieve these goals, the Mars model hides the programming complexity of GPUs by: (i) use of a simple and conversant MapReduce interface, (ii) routinely managing subdividing tasks, (iii) managing the dissemination of data, and (iv) dealing with process parallelization [9]. On the other hand, for efficient utilization of resources, Stuart and Owens (2011), proposed GPMR, a stand-alone MapReduce library that takes advantage of the power of GPU clusters for extensive computing by merging bulky quantities of map and reduce items into slices and using fractional reductions and accruals to complete computations [10].

Although the Pthreads model has been around for some time [11], application of the MapReduce model using Pthreads is accomplished via the Phoenix model – a programming paradigm that is built on Pthreads for MapReduce implementation [12]. Of recent, the Phoenix++ MapReduce model, a variation of Phoenix model, was created based on the C++ programming prototype [13]. Yet still, on the issue of determinism in Pthreads, Liu, Curtsinger, and Berger (2011), observed that enforcing determinism in threads was still problematic and they proposed the DTHREADS multi-threading model, built on C/C++ to replace the Pthreads library and alleviate the non-determinism problem [14].

DTHREADS model imposes determinism by separating multi-thread programs into several processes; using private and copy-on-write maps to shared memory, employing standard virtual memory protection to keep track of writes, and then deterministically ordering updates on each thread, and as such, eliminating false sharing [14]. However, the issue of multi-thread non-determinism in parallel programming models still persists, and continues to be problematic due to the data race issue, a condition brought about by unneeded shared memory communication between threads [15]. To address this problem Lu, Zhou, Wang, Zhang, and Li (2013), proposed RaceFree, a relaxed deterministic model that offers a data race free multi-threading setting whereby determinism in parallel programs is only affected by synchronization race [15].

## 2.1 HOW GPUs WORK

In this section a brief review of how a GPU pipeline work is given to help the programmer better understand and fully utilize the resources provided by parallelism in GPUs. In a tutorial on GPU architecture, Crawfis (2007) noted that a major difference between GPUs and CPUs, as illustrated in Fig 1, is that programs are executed in parallel in GPUs while the programs are executed serially in CPUs [16].

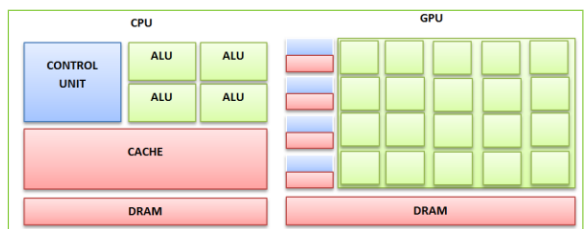

**Figure 1**. GPU has more ALUs than the CPU

Furthermore, Crawfis (2007) observed that another difference is that while CPUs have fewer executions units but higher clock speed, GPUs are composed of several parallel execution units, higher transistors counts, deeper pipelines, and faster advanced memory interfaces than CPUs. This makes GPUs much more robust, faster, and more computationally powerful than CPUs [16]. Both Crawfis (2007) and Lindholm, Nickolls, Oberman, and Montrym (2008) exemplified that the basic abstract structure of the GPU pipeline as pointed out in Fig 2, is composed of, (i) the host interface, (ii) the vertex processing, (iii) triangle setup, (iv) pixel processing, and (v) memory interface[16] [17].

*The host interface:* The host interface works as the communication link between the CPU and the GPU by responding to commands from the CPU, collecting data from the system memory, ensuring consistency in system commands, and performing context switching [16] [17]. *The vertex stage:* in this stage, as observed by both Crawfis (2007) and Owens et al., (2008), the vertex receives information from the host interface and outputs the vertices in screen space [16] [18]. *Triangle stage*: At this stage, geometry data is translated into raster information by translating screen space geometry information into raster pixel information, with each vertex computed in parallel, and output placed on the screen space [16] [18]. *The Fragment stage:* Triangle information is then used to compute final color for the pixels; it is at this stage that computation and math operations are done [16] [18].

*The memory interface stage*: in this stage, the color fragments generated in the fragment stage are written to the frame buffer [16]. Crawfis (2007) noted that one of the benefits of the GPU pipeline, is that vertex, fragment processing, and triangle setup are all programmable; giving programmers the ability to write programs that are executed for every vertex and fragment [16]. In addition, Owens, et al (2008) observed that the GPU pipeline helps attain parallelism for an application. In this way, data in multiple pipeline stages can be computed concurrently, and as such, achieve data parallelism, a crucial feature that programmers can exploit with GPUs [18].

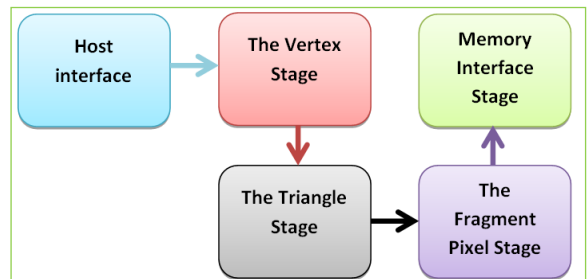

**Figure 2**: The GPU pipeline process.

## 3. CUDA PROGRAMMING FEATURES

CUDA also known as Compute Unified Device Architecture was developed in 2006 by NVIDIA as a general purpose parallel computing programming model, to run on NVIDIA GPUs to for parallel computations [19]. With CUDA, Programmers are granted access to GPU memory and therefore, are able to utilize parallel computation not only for graphic application but general purpose processing (GPGPU) [20]. One of the challenges of HPCs is how to fully take the parallelism advantage presented by the multi-core CPUs and many-core GPUs; CUDA programming language is designed to surmount this challenge by taking gain of parallelization in both CPUs and GPUs [19].

In this section, we take a look at some of the programming features provided by CUDA. While a number of features are made available by CUDA to the programmer, in this paper, we focus on features related to threads. *Process flow*: A CUDA program execution, as shown in Fig 3, is done in two parts, on the host – also known as the CPU, and on the device – also referred to as the GPU [21] [22]. CUDA programs interact with both the CPU and GPU during program execution. The process flow of a CUDA a program is then accomplished in the following three steps: (i) input data is copied from CPU memory to GPU memory; (ii) the GPU program is then loaded and executed; (iii) finally, results are copied from the GPU memory to the CPU memory [21] [22].

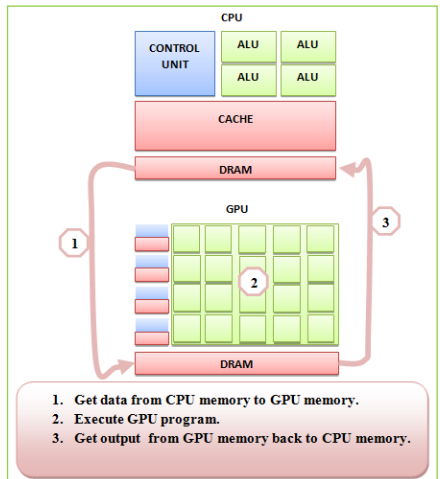

**Figure 3:** The CUDA process flow

*The kernel*: The basic feature of a CUDA program is the kernel. This is the heart of the GPU program. Programmers using CUDA, a language based on the C programming language, can generate functions called kernels that when called, get executed in parallel by different CUDA threads [19] [22]**.** To define a kernel in CUDA, the __global__ declaring function is used; to specify the number of threads, the <<<...>>> syntax is utilized [1] [22]. *A thread:* the smallest feature in the CUDA program model is a thread, as illustrated in Fig 4. In CUDA, a kernel is a form of C program single distinct thread that depicts how that thread does computation [17]. *A thread block*: Another feature that CUDA provides to programmers is the ability to group a batch of threads into blocks. A thread block, as shown in Fig 4, is a group of threads that get synchronized using barriers and communicate using shared memory [23].

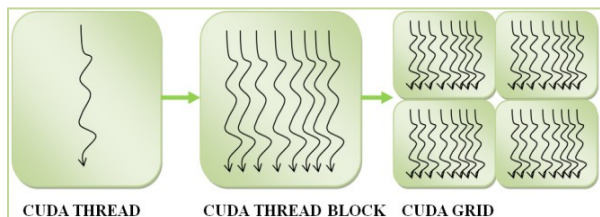

**Figure 4:** The CUDA thread, thread block, and grid

*The grid*: the grid, in CUDA, as demonstrated in Fig 4, is a group of thread blocks that can coordinate using atomic operations in a global memory space shared by all threads [23]. Threads in the grid get synchronized by means of global barriers and coordinate using global shared memory [23]. *Function types qualifiers:* According to the CUDA manual, CUDA provides function type qualifiers that specify if a function gets executed on the device (GPU) or host (CPU) and likewise if the function can be called from a device or host [19]. Fig 5, illustrates function types qualifiers used in the vector addition example. The function type qualifiers used in CUDA include [19]: *(i)__global__* denotes a kernel function that gets called on host and executed on device. *(ii)__device__* denotes device function that gets called and executed on device. *(iii)__host__* denotes a host function that gets called and executed on host. *(iv)__constant__* denotes a constant device variable that is accessible by all threads. *(v)__shared__* denotes a shared device variable available to all threads in a block.

*Data types*: CUDA provides built-in data types also referred to as vector types that are derived from the basic C language data types. These include char, short, int, long, long, float, and double. The vector types are structures that are accessible through the component values x, y, z, and w, using a constructor function structured as follows: make_<type name>; An example could include, float1 make_float1(float x, float y), and as further presented in Fig 6 [20] [24]. A comprehensive list of data types as provided by the CUDA programming guide includes [19] [24]: *(i) Characters:* char1, uchar1, char2, uchar2, char3, uchar3, char4, and uchar4. *(ii) Short:* short1, ushort1, short2, ushort2, short3, ushort3, short4, and ushort4. *(iii) Integers:* int1, uint1, int2, uint2, int3, uint3, int4, and the uint4. *(iv) Long:* long1, ulong1, long2, ulong2, long3, ulong3, long4, and ulong4. *(v) Long long*: longlong1, ulonglong1, longlong2, and ulonglong2. *(vi) Float:* float1, float2, float3, and float4. *(vii) Double*: ouble1 and double2

```
// Adding Vectors in CUDA

#include <stdio.h>
#include <stdlib.h>
#include <math.h>

// Function executed on device (GPU), gets called on Host(CPU)
__global__ void vectorAdd (double *a, double *b, double *c, int n)
{
    // Get block index, thread index, and dimension of block in threads
    int id = blockIdx.x * blockDim.x + threadIdx.x;

    // ensure array size is kept
    if (id < n)
        c[id] = a[id] + b[id];
}
```

**Figure 5:** Vector addition example in CUDA [25].

```
// Adding Vectors in CUDA

int main( int argc, char* argv[] )
{
    // Size of vectors
    int n = 20000;

    // Host (CPU) input vectors
    double *host_a;
    double * host_b;

    //Host (CPU) output vector
    double *host_c;

    // Device (GPU) input vectors
    double *device_a;
    double *device_b;
```

**Figure 6:** Vector addition: data types.

```
// Assign memory to every vector on host (CPU)
host_a = (double*)malloc(bytes);
host_b = (double*)malloc(bytes);
host_c = (double*)malloc(bytes);

// Assign memory to every vector on device (GPU)
cudaMalloc(&device_a, bytes);
cudaMalloc(&device_b, bytes);
cudaMalloc(&device_c, bytes);

int i;
// Assign the initial values and fill vectors on host (CPU)
for( i = 0; i < n; i++ ) {
    host_a[i] = sin(i)*sin(i);
    host_b[i] = cos(i)*cos(i);
}
```

**Figure 7:** Vector addition: memory allocation.

*Built-in variables*: CUDA comes with built-in variables that indicate the grid and block sizes (see Fig 8), and the block and thread indices, that are only applicable within a function and executed on the GPU; the variables include [19] [24]: *(i) gridDim* – denotes the dimensions of grid in blocks. *(ii) blockDim* – denotes the dimensions of block in threads. *(iii) blockIdx* – denotes a block index within grid. *(iv) threadIdx* – denotes a thread index within block.

*Thread management*: While determinism is difficult to attain in multi-threading, CUDA provides a number of threads management functions that provide determinism supervision [20] [24]: *(i) __threadfence_block()* – enforces a wait until memory is available to the thread block. *(ii) __threadfence()* – implements a wait until memory is accessible to a thread block and device. *(iii) __threadfence_system()* – imposes a wait until memory is available a block, device and host. *(iv) __syncthreads()* – enforces a wait until all threads coordinate through synchronization.

```
// Get host (CPU) vectors to device (GPU)
cudaMemcpy( device_a, host_a, bytes, cudaMemcpyHostToDevice);
cudaMemcpy( device_b, host_b, bytes, cudaMemcpyHostToDevice);

int blockSize, gridSize;

// Assign the number of threads for each thread block
blockSize = 1024;

// Assign the number of thread blocks in the grid
gridSize = (int)ceil((float)n/blockSize);

// Execute computation on the GPU
vectorAdd<<<gridSize, blockSize>>>(device_a, device_b, device_c, n);

// Get computation results from device (GPU) back to host (CPU)
cudaMemcpy( host_c, device_c, bytes, cudaMemcpyDeviceToHost );
```

**Figure 8:** Vector addition: vectors from CPU to GPU.

*Memory management:* A CUDA program is always hosted on the CPU while the computation gets done on the GPU, (see Fig 7 and Fig 9). The results are then sent back to the CPU, and as such, CUDA avails programmers with memory management tools that allocate and free memory on both host and device [20] [24]: *(i) cudaMalloc( )* – allocates memory on device. *(ii) cudaFree( )* – frees allocated memory on device. *(iii) cudaMemcpyHostToDevice, cudaMemcpy( )* – copies from host memory to device. *(iv) cudaMemcpyDeviceToHost, cudaMemcpy( )* – copies device results back to host memory.

*Arithmetic functions:* in the CUDA programmer's guide, provided by NVIDIA, the arithmetic functions made available in CUDA basically reads the 32-bit or 64-bit word *X* located at the address *Y* in global or shared memory, then do the computation, and store the result back to memory at the same address *Y*. The following are some of the arithmetic functions provided by CUDA [19]: *(i) atomicAdd ( )* – computes addition. *(ii) atomicSub ( )* – computes subtraction. *(iii) atomicExch ( )* – exchanges values. *(iii) atomicMin ( )* – computes the minimum value. *(iv) atomicMax ()* – computes the maximum value.

```
// Get the sum of results in vector c; divide sum by n, to equal 1 within error
double sum = 0;
for(i=0; i<n; i++)
    sum += host_c[i];
printf("final result: %f\n", sum/n);

// Free device (GPU) memory
cudaFree(device_a);
cudaFree(device_b);
cudaFree(device_c);

// Free host (GPU) memory
free(host_a);
free(host_b);
free(host_c);
 return 0;
}
```

**Figure 9:** Vector addition: freeing memory.

### 3.1 GPU VERSUS CPU EXPERIMENT

For illustration purposes, we implemented a vector addition program using CUDA on a HPC system with GPU and CPU processors. The goal of the demonstration was to show that GPUs are faster than CPUs, given an increase in workload – the number of integers in the vectors to be added. The experiment was run on a Cray XK7 HPC system with (i) 1856 AMD processor cores, (ii) about 50 TFLOPS, (iii) 2.4 TB memory, (iv) State-of-the art Gemini 2D torus interconnect, and (v) 32 latest NVIDIA Kepler K20 GPUs [25].

Table 1. Vector addition processing time

| Data size | GPU Time | CPU Time |
|---|---|---|
| 5 | 0.042 | 0.004 |
| 50 | 0.042 | 0.004 |
| 500 | 0.042 | 0.008 |
| 5000 | 0.042 | 0.064 |
| 50000 | 0.043 | 0.582 |
| 500000 | 0.043 | 6.32 |
| 5000000 | 0.055 | 64.601 |
| 50000000 | 0.043 | 640.364 |
| 500000000 | 0.041 | 6194.293 |

Table 1, shows the data size of integers in each vector to be added, ranging from 5 to 500 million integers. The GPU and CPU processing times also shown in Table 1, depict how long it took to complete the calculations in seconds. In Fig. 10, the CPU processing time remained constant, at an average time of 0.04 seconds, up until the load size was 5000 integers. At 50000 integers, there was an exponential rise in CPU processing time, from 0.064 seconds to 6194 seconds, even as the load size increased. In Fig 11, GPU processing time stays constant for much of the load size, with an average of 0.042 seconds. However, there is a small rise in GPU processing time when the load size is five million integers, with GPU processing time at 0.055 seconds. Yet still, this small rise in GPU processing time is far better than the CPU processing time for the same load size, at 64.601 seconds.

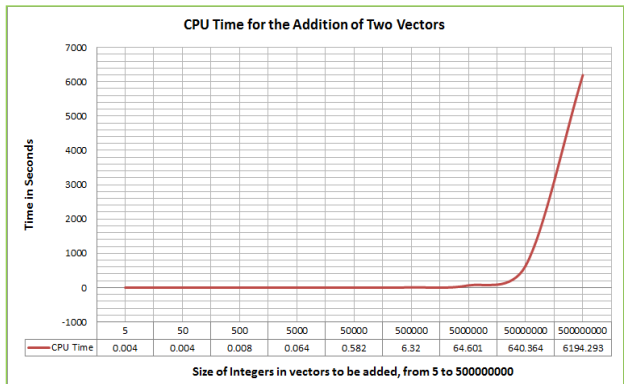

Figure 10. CPU processing time for the addition of two vectors

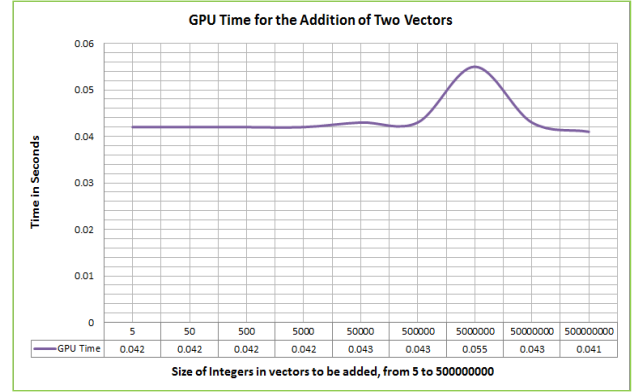

Figure 11. GPU processing time for the addition of two vectors

As the load size increases to 50 million integers, the GPU processing time falls back to the average of 0.042 seconds, at shown in Fig 11, and significantly outperforming the CPU processing time for the same load size at, 640.364 seconds. Therefore, GPUs outperform CPUs as the load size increases.

## 4. MAPREDUCE

There are many different ways to implement MapReduce and the correct manner depends of the environment that the program will be working in. MapReduce jobs have been created on NUMA multiprocessors, large connection of commodity nodes on a cluster, and shared memory machine models [33]. The functional mechanisms of the MapReduce programming model takes a set of key/value input pairs and then reduces to another set of output key/value pairs. The user of the routine library can express this within a map or reduce function. The map function creates a mapping of all key/value in the form of intermediate key/value pairs. The MapReduce library automatically groups all intermediate key/value pairs together and passes them to the reduce function. The reduce function takes an argument of the intermediate key/value pairs. For intermediate values, the argument is a key/value pair that has many values e.g. <key, value1, value2, …, value n> [33]. The reduce function then merges together these values to form a smaller set of values. The values are supplied to the reduce function by the MapReduce iterator which allows the function to handle large lists of values as input in order to fit in memory. The map and reduce functions produce results from two different domains, furthermore the intermediate values allows these domains to be migrated as follows:

*map  (k1,v1) → list(k2,v2)*

*reduce (k2,list(v2)) → list(v2)*

```
map(String key, String value)
        //key: document name
        //value: document contents
        for each word w in value:
                EmitIntermediate(w, "1");

reduce(String key, Iterator value);
        //key: a word
        //values: a list of counts
        int result = 0;
        for each v in values:
                result+=ParseInt(v);
        Emit(AsString(result));
```

**Figure 12.** Pseudocode of MapReduce word count [33].

In Fig 12, the MapReduce pseudocode in an implementation of the programming model which strives to succumb to the issue of counting the number of times a word occurs in a document or a set of documents. The map function creates a key/value pair of the name of the word and the initial occurrence of each word which will be 1. The reduce function then sums all the counts that were created by the mapper. There is a MapReduce specification object that takes in arguments of the names of the input and output files associated with the initial data to be computed. Figure. 13, gives an execution overview of MapReduce and how the programming model has a workflow that initializes a user program which forks tasks to different workers. A Master has the ability to control the flow of forking different tasks/jobs to associated workers. This allows the master to be in control of the designation jobs to different workers and the corrective planning of inevitable fault tolerance issues associated with node failure. The initial data is split or distributed across the nodes on the network/cluster using a unique file system structure [35].

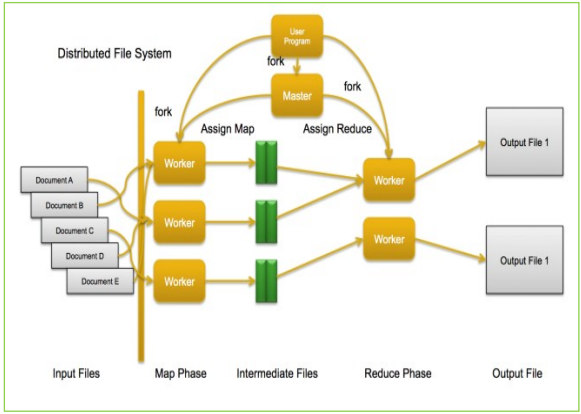

**Figure 13.** MapReduce execution overview [33].

The master designates jobs to workers that must find the data associated with their job on the host node, and execute the associate read/write task. The intermediate files are created on the local disks by the map phase. The reduce phase now takes over in which the Master will assign a set of workers to complete the reduce task/jobs. The reduce phase now creates a smaller subset of key/value pairs to be merged together and results sent to the data store [34].

```
import java.io.IOException;
import java.util.*;
import org.apache.hadoop.fs.Path;
import org.apache.hadoop.conf.*;
import org.apache.hadoop.io.*;
import org.apache.hadoop.mapreduce.*;
import org.apache.hadoop.mapreduce.lib.input.FileInputFormat;
import org.apache.hadoop.mapreduce.lib.input.TextInputFormat;
import org.apache.hadoop.mapreduce.lib.output.FileOutputFormat;
import org.apache.hadoop.mapreduce.lib.output.TextOutputFormat;

public class WordCount {
public static class Map extends Mapper<LongWritable, Text, Text, IntWritable> {
private final static IntWritable one = new IntWritable(1);
private Text word = new Text();

public void map(LongWritable key, Text value, Context context) throws IOException, InterruptedException {
String line = value.toString();
StringTokenizer tokenizer = new StringTokenizer(line);
while (tokenizer.hasMoreTokens()) {
word.set(tokenizer.nextToken());
context.write(word, one);
                }
        }
```

**Figure 14.** MapReduce initialization and word count map function [33].

Figure 14 introduces the MapReduce programming environment and framework. The code demonstrated how MapReduce is initialized through the use of its Java API. The API contains function call in order to implement MapReduce through distribution, scaling, and parallization with some of the following attributes: (i) Job Configuration, (ii) Task Execution & Environment, (iii) Memory Management, (iv) Map Parameters, (v) Shuffle/Reduce Parameters, (vi) Directory Structure, (vii) Task JVM Reuse, (viii) Configured Parameters, (ix) Task Logs, and (x) Distributing Libraries. Figures 15 introduces the MapReduce's map and reduce function and how it is used in a low commodity cluster environment. The MapReduce Java API uses a write call to write the results from the map phase to intermediate key/value pairs that hold a <key, word occurrence> for each word associated in a document.

```
public static class Reduce extends Reducer <Text, IntWritable, Text, IntWritable>
    {
        public void reduce(Text key, Iterator<IntWritable> values, Context context)
            throws IOException, InterruptedException {
                int sum = 0;
                while (values.hasNext()) {
                    sum += values.next().get();
                }
                context.write(key, new IntWritable(sum));
            }
    }
public static void main(String[] args) throws Exception {
    Configuration conf = new Configuration();
    Job job = new Job(conf, "wordcount");
    job.setOutputKeyClass(Text.class);
    job.setOutputValueClass(IntWritable.class);
    job.setMapperClass(Map.class);
    job.setReducerClass(Reduce.class);
    job.setInputFormatClass(TextInputFormat.class);
    job.setOutputFormatClass(TextOutputFormat.class);
    FileInputFormat.addInputPath(job, new Path(args[0]));
    FileOutputFormat.setOutputPath(job, new Path(args[1]));
    job.waitForCompletion(true);
        }
    }
```

**Figure 15.** MapReduce word count reduce function [33].

The reduce phase then use the API function write to compute a new pair of key/value pairs that now that the sum of the total occurrences of each word in the form of <key, total word occurrence>.

```
Assuming HADOOP_HOME is the root of the installation and HADOOP_VERSION is
the Hadoop version installed,

compile WordCount.java and create a jar:
    $mkdir wordcount_classes
        $ javac -classpath ${HADOOP_HOME}/hadoop-${HADOOP_VERSION}-core.jar
        -d wordcount_classes WordCount.java
        $ jar -cvf /usr/joe/wordcount.jar -C wordcount_classes/ .

Assuming that:
        /usr/joe/wordcount/input - input directory in HDFS
        /usr/joe/wordcount/output - output directory in HDFS

Sample text-files as input:
        $ bin/hadoop dfs -ls /usr/joe/wordcount/input/
        /usr/joe/wordcount/input/file01
        /usr/joe/wordcount/input/file02

        $ bin/hadoop dfs -cat /usr/joe/wordcount/input/file01
Hello World Bye World
        $ bin/hadoop dfs -cat /usr/joe/wordcount/input/file02
Hello Hadoop Goodbye Hadoop

Run the application:
        $ bin/hadoop jar /usr/joe/wordcount.jar org.myorg.WordCount
/usr/joe/wordcount/input /usr/joe/wordcount/output
Output:
        $ bin/hadoop dfs -cat /usr/joe/wordcount/output/part-00000
Bye 1 Goodbye 1 Hadoop 2  Hello 2  World 2
```
**Figure 16.** MapReduce command line execution overview [33]

Figure 16 depicts the command line arguments needed to compile the java word count file. The input files that are to be parsed in order to count the words in the document are to be stored using the associated file system structured mentioned by Ghemawat, et al. [35]. In order for MapReduce to be successfully compiled, an open source version of Google File System, MapReduce, and Bigtable need to be installed. For this example Apache's Hadoop Distributed File System (HDFS), Hadoop Base, and Hadoop MapReduce was installed. The input file was stored using HDFS.

## 5. PTHREADS

The first POSIX (portable operating system interface) threads model was created by the IEEE computer society in 1988 and current versions are maintained by the Austin common standards revision group [28] [30]. One very popular application programming interface (API) for multi-threading, is Pthreads, also known as POSIX threads, with the IEEE and ISO notation of P1003.1c and ISO/IEC 9945-1:1990c respectively [29] [27]. Pthreads programming model is part of the C programming language and composed of programming types and procedure calls, that get called and included by a *pthread.h* header thread library; however, the *pthread.h* header library might be part of another library in other Pthreads implementations [29] [27] [30]. Pthreads do not depend on data transfer but utilize the cache to CPU and memory to CPU bandwidth for data transfer, thus making Pthreads process computations much faster [27]. To fully exploit the computational potential presented by Pthreads, a program has to be structured into distinct autonomous processes which then get implemented and run in parallel on a high performance machine [27]. An example would include sub-routine1 and sub-routine 2 being swapped, enclosed, and overlaid in real time, thus making them suitable for multi-threading [27].

Pthreads are composed of about 100 Pthreads sub-routines, which begin with a "*pthread_*" preface and are largely classified as follows [29] [27] [32]: *(i) Thread management sub-routines*: these are sub-routines that affect threads directly, for example creating, detaching, and joining. *(ii) Mutexes sub-routines*: these are sub-routines that handle synchronization in threads, by using mutual exclusion (mutex) to enforce determinism by harmonization in threads. *(iii) Condition variable sub-routines*: these are sub-routines that handle communication between threads that have a common mutex in enforcing synchronization. *(iv) Synchronization sub-routines*: these are sub-routines that enforce determinism in threads by managing read and write locks, and barriers. One of the drawbacks of Pthreads is how to manage the persistent problem of non-determinism [30]. POSIX threads depend on thread mutexes, condition variables, and synchronization control methods in managing determinism in threads; with mutexes only permitting one thread to enter a critical section at a time and avoiding deadlocks [1].

The following are some of the most utilized Pthreads prefixed sub-routines [27]: *(i) pthread_:* denotes the threads used and other various sub-routines. *(ii) pthread_attr_:* denotes the thread attributes. *(iii) pthread_mutex_:* denotes the mutex functionality for threads. *(iv) pthread_mutexattr_:* denotes the mutex attributes. *(v) pthread_cond_:* denotes the condition variables in threads. *(v) pthread_condattr_:* denotes the condition attributes. *(vi) pthread_key_:* denotes the data keys for threads. *(vii) pthread_rwlock_:* denotes read and write locks utilized in managing determinism in threads. *(viii) pthread_barrier_:* denotes the synchronization function used to manage determinism in threads.

```
//Create
#include <pthread.h>
int
pthread_create
(pthread_t *thread_id, const pthread_attr_t *attributes,
  void *(*thread_function)(void *), void *arguments);

//Waiting
int
pthread_cond_wait (pthread_cond_t *cond, pthread_mutex_t *mut);

//Waiting with timeout
int
pthread_cond_timedwait (pthread_cond_t *cond, pthread_mutex_t *mut,
                       const struct timespec *abstime);

//Terminating threads
int
pthread_exit (void *status);
```

**Figure 17:** Pthread creation, waiting, and termination [31].

Generating and discarding threads: Pthreads is composed of two main functions responsible for creating and terminating threads, namely, *pthread_create* and *pthread_exit* [30] [1] [27].

```
//Pthreads Mutex Example
THREAD 1                              THREAD 2

pthread_mutex_lock (&mut);
                                      pthread_mutex_lock (&mut);
a = data;                             /* blocked */
a++;                                  /* blocked */
data = a;                             /* blocked */
pthread_mutex_unlock (&mut);          /* blocked */
                                      b = data;
                                      data = b;
                                      b--;
                                      pthread_mutex_unlock (&mut);
```

**Figure 18:** Pthreads Mutex example [31].

The *pthread_create* function constrains a programmer to stipulate the thread used to run tasks, the attribute, tasks to be run by thread in a routine call, and the routine argument [30] [1]. The *pthread_exit* function permits the programmer to stipulate how and when a thread gets discarded, by either returning normally from its subroutine when its work is done, or having the thread terminated by another thread [30] [27] [1].

## 6. CONCLUSION

In this paper, we have endeavored to present a succinct overview of three parallel programming models, CUDA, MapReduce, and Pthreads. The goal is to assist high performance users with a concise understanding of parallel programming concepts and thus faster implementation of big data projects using high performance computing. Although there are many parallel, distributed, scalable programming that currently in existence, there was no model whose main goal through development was to create a high performance, distributed, parallel, and scalable programming environment that was easy to use. The innate ability of MapReduce to do it's parallel and distributed computation across large commodity clusters at run-time allowed developers to express simple computations while hiding the messy details of communication, parallelization, fault tolerance, and load balancing. Furthermore, the use of job distribution through map and reduce phases, while using a distributed file system architecture for data storage and retrieval to produce a framework for easily writing applications which process vast amounts of data in parallel on large clusters using commodity hardware in a fault tolerant manner. Future works will focus on how MapReduce could be implemented in both CUDA and Pthreads.